
\input phyzzm.tex

\theory
\pubnum{29}
\pubtype{}

\titlepage
\title{ANTIMATTER IN THE UNIVERSE\foot{Invited talk at "Antihydrogen
Workshop", Munich, July, 1992.}}
\author {A. D. Dolgov}
\address {Randall Physics Laboratory\break
University of Michgan\break
Ann Arbor, MI 48109-1120\break
U.S.A.}

\abstract
\smallskip

Cosmological models which predict a large amount of antimatter in the
Universe are reviewed. Observational signatures and searches for cosmic
antimatter are briefly considered. A short discussion of new long range
forces which might be associated with matter and antimatter is presented.

\endpage

\centerline {{\bf 1.Introduction.}}

The world in our neighbourhood, as we observe it, is 100\% charge asymmetric.
We see only matter and any considerable amount of antimatter nearby
is excluded [1,2].
A small number of antiprotons and positrons in the cosmic rays
(to say nothing of those produced in accelerators) can be prescribed to the
secondary origin. Charge asymmetry in term of {\it visible} massive matter
defined as
$\beta_B = (n_B -n_{\bar B} ) / (n_B + n_{\bar B}) $
is close to one. Here $n_B$ and $n_{\bar B}$ are number density of
baryons and antibaryons respectively. However the cosmologically more
natural quantity
$$
\beta ={n_B -n_{\bar B} \over n_\gamma} \approx 3\cdot 10^{-10}
\eqno (1)
$$
is surprisingly small. Here $n_\gamma \approx 400/cm^3 $ is the number
density of photons in electromagnetic background radiation with temperature
$T=2.7$K. The smallness of $\beta$ implies that $\beta_B$ was also small
and of the same order of magnitude as $\beta$ at an early stage of
the Universe evolution when the temperature was about or above the proton
mass and antibaryons were abundantly produced in the primeval plasma.
(To be more exact the characteristic temperature is that of the QCD
phase transition, $T_{QCD} \approx 100$ MeV, when free quarks are confined
to hadrons.)

At the present day we are able to directly observe only a  minor fraction
(of order of a few per cent) of the total mass of the Universe. Nothing
is known about the physical nature of the dark matter and in particular
if it is charge asymmetric or  not. If the dark matter consists of massive
($m=O(10{\rm eV})$) neutrinos then most naturally the neutrino asymmetry
$\beta _\nu = (n_\nu -n_{\bar \nu})/ (n_\nu + n_{\bar \nu} )$ is
tiny, $\beta _\nu = 10^{-9} - 10^{-10}$, though $\beta_\nu = O(1)$
is not excluded. Astronomical data disfavor very much neutrino dark matter
hypothesis and other candidates are being looked for. If the latter is
neutral particle (like Majorana type neutralino in supersymmetric models
or axion) the charge asymmetry in the dark matter sector is zero by
definition. In other cases nothing is known about charge asymmetry of
dark matter.

In this talk I will address the following questions

\item {1.} Why there is a charge asymmetry in the Universe?
\item {2.} Why it is so small?
\item {3.} May the Universe be charge asymmetric locally and charge
   symmetric globally?
\item {4.} What would be the size of domains with a definite sign of the
   charge asymmetry?
\item {5.} What are possibilities of observation of antimatter?
\item {6.} Can there be any difference in gravitational interactions of
   matter and antimatter and can be there new long range forces different
   for matter and antimatter?

This talk is not intended to be a review on cosmology though a knowledge of
some cosmological material is desirable. It would be hard to give all the
appropriate references here in this short talk and a reader can find them
as well as necessary detail in numerous review papers and textbooks.

\bigskip

\centerline {{\bf 2. Baryogenesis.}}
\bigskip

The origin of the baryon asymmetry was understood after the paper
by Sakharov [3] where three basic principles of baryogenesis were formulated.
They are:

\item {1.} Breaking of charge (C) and  combined charge-parity (CP)
invariance.
\item {2.} Deviation from thermal equilibrium in the primeval plasma.
\item {3.} Baryonic charge (B) nonconservation.

It was shown that under these conditions a charge (baryonic) asymmetry
is developed from practically arbitrary initial state.

These three principles stay on rather different footing. Thermal
nonequilibrium is induced by the nonabiabaticity created by the Universe
expansion. This is true for massive particles but usually is rather small
if some special conditions are not realized. The breaking of C and CP
is well established experimentally but not well understood theoretically
and not much is known about charge asymmetry breaking at high energies
at which baryonic charge is presumably nonconserved. A nonconservation of
the latter was not observed in direct experiments despite hard attempts to
discover the proton decay. Theoretically it looks at least very natural
not to say strictly proven. First, the models of grand unification which
put quarks and leptons into the same particle multiplet (like $SU(5)$ [7])
predict transition between quarks and leptons with corresponding
nonconservation of baryonic and leptonic charges. The characteristic
energy scale of such processes is huge, $m_{GUT} = 10^{15} -10^{16}$ GeV
so in these models baryogenesis took place at very high temperatures
very close to the initial singular state of the Universe.

A more interesting possibility is the baryonic charge nonconservation
in the standard electroweak $SU(2) \times U(1)$ theory [8]. This
phenomenon is really striking because the classical Lagrangian conserves
baryonic charge and the nonconservation arises as a result of quantum
corrections and proceeds as some kind of quantum tunnelling process.

Thus baryonic charge nonconservation is predicted theoretically though
we still lack experimental confirmation. The only evidence in favor of
B-nonconservation is given by cosmology. This is not only the baryon
asymmetry itself but also an impossibility to realize inflationary
universe scenario if B is conserved. Since at the present time inflation
gives the only way to create the observed Universe and in that sense can be
considered as an experimental fact, one may think that baryon
nonconservation is also proven experimentally. Though it is possible to
invent some counterexamples to this statement (see review paper [6]),
they seem to be much more exotic than baryon nonconservation.

Majority of the baryogenesis scenarios give a very low value of $\beta$
and the special efforts are taken to get the observed value (1). The
reasons for getting a small $\beta$ are the following. First, even if
the CP-violating amplitudes are not suppressed with respect to CP-even
ones still their observational manifestations are rather weak and arise
only in a high order in a small coupling constant because CP-violation
becomes observable only through interference effects with CP-conserving
processes. Second, the deviation from thermal equilibrium in the early
Universe are typically small because the expansion is mostly slow in
comparison with the characteristic reaction rates. And last, but not the
least, there is a danger of a large entropy generation after baryogenesis
(by the first order phase transition or by heavy particle annihilation)
which can considerably dilute a previously generated charge asymmetry.
There is only one known to me example [10] when the calculated value
of the baryon asymmetry happened to be (much) larger than the observed
value (1) and one had to invent a way to dilute it by the entropy
production in the course of the Universe expansion and cooling down.

In short the model of ref. [10] is the following. It is assumed that
there exists a scalar field $\chi$ with a nonzero baryonic charge.
Such fields naturally appear in supersymmetric theories. They are
scalar superpartners of quarks. During inflationary stage all the
matter fields except for scalar ones exponentially die out while
scalar fields are even amplified due to rising quantum fluctuation if
their mass is smaller than the Hubble parameter, $m_\chi <H_I$ [11].
If this is true for $\chi$ one would expect that a condensate of
baryonic charge (stored in $\chi$) should be developed during during
inflation with a large (generically of the order of $H_I^3$) baryonic
charge density. When inflation is over and $\chi$ relaxes to the
equilibrium state, its coherent oscillations produce an  excess of
quarks over antiquarks or vice versa depending upon the initial
sign of the baryonic charge condensate. The latter is generated
stochastically and is zero if averaged over a large volume. In this
model the Universe is charge symmetric as a whole while it might have
a significant charge asymmetry locally. Note that this scenario is
effective even without CP-violation in the Lagrangian. In a more
traditional approach the similar Universe structure appears in the
models with spontaneous CP-violation [12].

\bigskip

\centerline {{\bf 3. CP-violation and a charge asymmetry of the Universe.}}
\bigskip

If the symmetry between particles and antiparticles is broken explicitly
in the Lagrangian the charge asymmetry should have the same sign all over
the  Universe. There might be however counterexamples to this statement:
as was argued in ref. [13] in some models the sign of the asymmetry may
be different depending upon the rate of the cooling. So if the cooling
was not homogeneous there could be domain of matter and antimatter in the
Universe even if CP is broken explicitly. The effect is much more profound
however in the models with spontaneous symmetry breaking. In this case there
should be an equal amount of matter and antimatter in the Universe.

Spontaneous symmetry breaking means that the Lagrangian is invariant with
respect to the symmetry transformation while the vacuum state is not. Still
the symmetry manifests itself by the presence of several (or infinitely many)
degenerate vacuum states related by the symmetry transformations.

Spontaneous CP-violation can be realized in the model with a complex scalar
field $\phi$ having the potential:
$$
U(\phi ) =-m^2 \mid \phi \mid ^2 +\lambda_1 \mid \phi \mid ^4 +
\lambda_2 ( \phi^4 + \phi^{\star 4} ) +g^2 T^2 \mid \phi \mid ^2
\eqno (2)
$$
The last temperature dependent term comes from the interaction with particle
in the thermal bath, $g$ being the coupling constant of the interaction.
One sees that at high temperatures the potential has the only minimum at
$\phi =0$ and CP is unbroken over this minimum. With decreasing temperature
two new minima arise while that at $\phi =0$ turns into maximum. The field
$\phi$ should evolve down to one of those two minima where it would have
a nonzero vacuum expectation value. A nonzero complex field condensate
results in the breaking of charge invariance. In particular through
Yukawa coupling to fermions it gives them complex masses which are equivalent
to complex angles in the Kobayashi-Maskawa matrix and correspondingly to CP
violation.

It is evident that the relaxation of $\phi$ to the minimum of the potential
would proceed with equal probability to the complex conjugate states and
so both sign of the CP-odd amplitude are equally probable. In this model
wee would expect domain of matter and antimatter equally probable in the
Universe [14].

Two problems arise in this case. First, the size of the domains should be
very small in the standard Friedman cosmology. Indeed the characteristic
size  of the domains at the moment of their formation should be smaller
than the horizon at this moment, $l\leq t_i$. After that, if it expanded as
the scale factor, it grew to the present time as $(T_i /2.7 {\rm K)}$ where
$T_i$ is the temperature at the moment of the formation and 2.7 K is the
temperature of the cosmic microwave radiation now. $T_i$ should be
definitely larger than 1 GeV since there is no baryon nonconservation at
this small temperature. Since $t\approx m_{Pl} /T^2 $ (where $m_{Pl}=
1.22\times 10^{19} $ GeV is the Planck mass) the present-day size
of the domains, $l_B$,
is definitely smaller than 10 light years. To be more precise the domain
walls could move with the speed close to that of light but since this
motion is chaotic, the average increase of their size does not differ much
from the above given estimate.
This problem can be solved if there existed a period of exponential
expansion (inflation) $a(t) \sim \exp (Ht)$ which could inflate the
domains up to arbitrary large sizes [15]. As a result the domain  sizes
would be exponentially poorly known. They can be larger than the
present-day horizon and so we effectively return to the charge symmetric
Universe as in the models with explicit CP-violation. The size of the
domains might be much smaller than that and so we live in the Universe
with equal amount of matter and antimatter inside the present-day horizon.
Of course the domain size cannot be too small, otherwise the bright results
of $p \bar p$-annihilation would be observed. The absence of antiprotons in
cosmic rays implies the bound $l_B > 10$ Mpc.

Another problem connected with this model is the existence of  domain walls
separating different vacua. Usually they have a large mass and produce
unacceptable density perturbations [16]. For the solution of the latter
a low temperature inverse phase transition is necessary. Note that the
domain wall problem does not appear in the model of baryogenesis with
baryonic charge condensate discussed above.

Thus we see that there is a solid theoretical  framework permitting to expect
that the Universe contains an equal amount of matter and antimatter. In the
case of the domain size being smaller than the horizon we may hope to
observe antimatter in the Universe.

\bigskip

\centerline {{\bf 4. More exotic models with
abundant antimatter in the Universe.}}
\bigskip

The period of the exponential expansion of the Universe together with
the CP-violation induced by a scalar field condensate may result in
a peculiar pattern of matter-antimatter distribution in the Universe.
To start with we consider a model which gives rise to a strictly periodic
distribution of baryons with possible alternation of baryonic and
antibaryonic layers [17]. Although it looks very much exotic the scenario
can be realized with a few very natural assumptions: first, the existence
of a complex scalar field $\phi$ with the mass which is smaller than the
Hubble parameter at the inflationary stage, $m_\phi < H_I$. Second,
nonharmonic terms should be present in the potential of $\phi$ like
e.g. $\lambda \mid \phi \mid ^4 $. Third, a condensate of $\phi$
should be developed during inflation which is not spatially constant
but a slowly varying function of space, $\phi (r) $. The first two
assumptions are perfectly natural, while the third one, though natural
too, deserves some explanatory remarks. A scalar field condensate is
generically formed in the De Sitter background due to rising quantum
fluctuations of $\phi$ [11] which produce spatially nonuniform but
effectively
classical field $\phi (r)$. Another possibility to get $\phi (r)$ is
a first order phase transition as result of which a bubble of $\phi $
is formed described by a function $\phi (r)$. The concrete form of
this function depends on the details of the phase transition but
what is certain their shape is not constant in space.
Once the bubble is formed  it
may remain frozen till inflation goes on and only when it is over the
field $\phi$ starts to oscillate  around the minimum of the potential.
If the baryogenesis proceeds prior to the complete relaxation
of $\phi$, the
space distribution of the baryons in this model should be periodic.
Indeed the size of the baryon asymmetry is proportional to the amplitude
of CP-violation and the latter is given by the complex field $\phi$.
Initially $\phi$ was a very slowly varying function of $r$ but in the
course of the evolution when $\phi$ quickly oscillates at each space
point as a function of time, a large wave periodic distribution  in
space would be developed. The latter is induced by nonharmonic terms
in the potential   because the period of oscillations in time depends
upon the initial amplitude. Such model naturally explains periodic
distribution of the visible matter observed in the recent pencil beam
measurements [18]. The distance between the layers of matter is about
100 Mpc and 5 layers were observed both in the direction to the North
and South galactic poles. If $\phi$ oscillates around zero (though it
is not obligatory) the baryonic layers should alternate with antibaryonic
ones separated by regions with a smaller matter density.

Another model [19] of abundant generation of antimatter in the Universe was
stimulated by an attempt to save the hypothesis of baryonic dark matter.
It is known that the bulk of matter in the Universe (from 90 to 99\%)
is invisible and is observed only by its gravitational action. What's
more the invisible (or dark matter) is most probably nonbaryonic.
There are several pieces of data in favor of these statements. The
velocity of gas around galaxies, $v$, does not decrease with the distance
from the luminous center but tends to a constant value (of the order of
few hundred km/sec). The observations have been done up to distances
almost order of magnitude larger than the size of galaxies. Such so called
flat rotational curves indicate that the mass is not concentrated in the
luminous central part but increases proportionally to the distance from
the center. The fraction of the invisible mass is different for different
galaxies but typically it is an order of magnitude larger than the visible
part. It is not easy to conceal all this matter if it is of the usual
baryonic staff. The option which is not excluded is that the dark matter
consists of large planets ("Jupiters"), or dead stars, or black holes
though the mechanism of their production is not clear.
There are however strong arguments against the hypothesis that all the
dark matter is baryonic. They are based on the theory of large scale
Universe structure formation and primordial nucleosynthesis. Large
scale Universe structure was formed from initially small perturbations
in the matter/energy density due to gravitational instability. The
growth of the perturbations in the baryonic matter could only start
rather late when the temperature dropped below that of the hydrogen
recombination, $T\approx 4000$ K. Above that temperature the light
pressure prevents electrically charged massive matter from gravitational
clumping. In the course of the expansion the inhomogeneities could rise
at most by the factor $z_{rec} = T_{rec} /T_{now} \approx 1.5\times
10^3$ where $T_{now}$ is the present-day temperature of the background
radiation. The initial (at $T=T_{rec}$ ) density inhomogeneities are
imprinted on the angular fluctuations of the background radiation
temperature. The quantity $\Delta T /T $ remains unchanged after the
recombination in the standard cosmological scenario and is known
from many observations at different angular scales to be
very small, $\Delta T/T \leq 10^{-5}$. Until recently there were
only upper bounds on $\Delta T/T$ and the first positive indication
for the nonzero value of it was presented only a few months ago
by COBE [20]. In view of that it is very hard (if possible) to get
developed density perturbations from the time of the recombination
to the present epoch. The existence of the dark (noninteracting
with light) matter makes things easier since the perturbations in
the latter may start earlier (before the recombination) when the
Universe became dominated by nonrelativistic (dark) matter.

Another argument against baryonic dark matter is based on a good agreement
of the calculated abundances of light elements,
created when the Universe
was only several minutes old, with observations (for a recent review on the
subject see e.g. ref. [21]). One of the parameters which determine the
production of light elements is the ratio of baryon-to-photon number
densities $\beta = n_B /n_{\gamma}$. Its value inferred from the data
on the light nuclei abundances is very close to that obtained by the
direct observation of the visible matter. There is some discrepancy
(like the factor of two) between these two values indicating
that some dark matter might be baryonic. Still the bulk of dark matter
is most probably nonbaryonic. It is convenient to characterize the
density of matter in the Universe $\rho$ by the parameter $\Omega =
\rho /\rho_c $ where $\rho_c$ is the closure  or critical energy
density given by the expression
$$
\rho_c =3H^2/8\pi G \approx 2\times 10^{-29} {\rm g/cm}^3 (H/100 {\rm
km/sec/Mpc} )^2
\eqno (3)$$
where $G$ is the Newton gravitational constant and $H$ is the present-day
value of the Hubble parameter. The latter is known with a rather bad
accuracy, $H=(50-100) $ km/sec/Mpc. For the visible baryonic matter
$\Omega ^{vis}_B \approx 10^{-2}$ which is close to that obtained from
the observed abundances of light elements (maybe a factor 2-3 smaller).
Analysis of the rotational curves gives $\Omega _{DM}^{clustered}
\approx 0.2-0.3$. This number refers to the matter clustered around
galaxies since the data are not sensitive to the uniformly distributed
matter. With the latter included $\Omega$ can be as large as 1 which is
strongly favored by the inflationary scenario. Since this scenario
presents the only known now way to resolve long-standing cosmological
problems, inflation can be considered as an experimental fact and we
have to conclude that 95-99\% of matter in the Universe is made of
some unknown  staff.

At the first glance the most natural candidate for the latter is massive
neutrino with $m=O(10{\rm eV})$. This particle is known to exist and we do not
see any reason why it should be massless. Unfortunately the Universe
structure in the model with massive neutrinos differs very much from
what is observed. The next possibilities the lightest supersymmetric
particle which most probably should be stable. These particles have
the generic name neutralinos and should be rather heavy, above a few
tens of GeV.

The observed picture supports the idea of two forms of dark matter, one
being clustered around galaxies (with $\Omega =0.2-0.3$) and another
uniformly distributed all over the Universe
(with $\Omega =0.7-0.8$). This might mean
that both light and heavy particles give a contribution into
dark matter, though it might be as well that the uniformly distributed
dark matter is the so-called cosmological term or in other words
the vacuum energy.

One more particle which is theoretically predicted and might provide
dark matter is the axion which ensures a natural CP-conservation in
quantum chromodynamics. Though very light ($m_a\leq 10 ^{-5}$ eV) the
axion by some funny reason behaves as heavy particles from the point of
structure formation creating the so-called cold dark matter, the same
type as lightest SUSY particles.

Dark matter in the Universe should not necessary be in the form of gas of
elementary particles but can be made of stable
macroscopical field configurations which maintain their stability by
topological reasons. These are so called topological solitons or
topological defects which arise in theories with spontaneously broken
symmetries. They include domain walls separating different degenerate
vacuum states, cosmic vorticies or strings, global monopoles, and
three dimensional textures. The latter are unstable but their life-time
is long enough to make them cosmologically interesting.

All these possibilities are certainly unusual (except for neutrino which
is almost for sure excluded) and demand a new physics beyond the standard
model. Supersymmetric particles may be in the best shape since supersymmetry
is a very appealing extension of the standard model. Still it is very
interesting to pursue the possibility of dark matter being made of usual
baryons (and as we shall see in what follows of antibaryons as well).
This can be realised with a special model of baryogenesis which gives
rise to relatively small bubbles of very high baryonic or antibaryonic
number density with $\mid \beta \mid =O(1)$ in the homogeneous baryonic
background with $\beta =10^{-9} -10^{-10}$ [19]. An interesting feature
of this model is that it predicts an equal number of baryonic and
antibaryonic bubbles so the Universe is practically charge symmetric.
At first sight such a picture strongly contradicts observations but
since the regions of high $\mid \beta \mid$ would mostly collapse into
black holes one can make the model compatible with the astronomical data.
In the limiting (and noninteresting) case of 100\% of these high density
baryon bubbles forming black holes there would be no observational
signatures characteristic to the model, except for the conclusion that
dark matter consists of black holes with masses about two orders of
magnitude around the Solar mass. A more exciting possibility is
that some of those bubbles did not collapse so that there are some
"naked" compact objects made from antimatter in the Universe.
Depending upon the relation between the Jeans wave length and the
horizon, at the moment when the latter is equal to the size of the
bubble, the bubble could form either a star (antistar) or a disperse
though well localized cloud of high density baryons or antibaryons.
They can be observed by fluxes of $\gamma$-rays from the annihilation
on their boundaries though it is hard to make a certain prediction
of their intensity. An unambiguous proof of existence of large amount
of antimatter would be an observation of antinuclei in cosmic rays.
The secondary production of the latter is practically impossible.
Another interesting phenomenon is a collision of a star with an
antistar. It would produce a powerful burst of gamma-radiation.
The observed by BATSE experiment on Gamma Ray Observatory isotropically
distributed $\gamma$-bursts are most probably explained by another
mechanism, though no completely satisfactory model is known at the
moment.
\bigskip
\centerline {{\bf 5. Search for cosmic antimatter.}}
\bigskip

There are three possible ways to detect the cosmic antimatter by
looking for antiprotons, antinuclei, and energetic ($\sim 100$ MeV)
gamma-quanta in cosmic rays. For the review of the possibilities of the
detection see refs. [22,23]. There were some data indicating an excess
of cosmic antiprotons over the theoretical expectations based on the
assumption of their secondary production. Unfortunately the most recent
and most accurate experiment [24] does not confirm the previous data
and gives only the upper bound $2\times 10^{-5}$ at 95\% C.L. for the
$\bar p /p $-ratio. This bound does not contradict the expectation
for the secondary produced $\bar p$ in the standard model as well as to
their production by cosmic neutralinos with $m>20$ GeV. The data of
ref. [24] together with the results of other experiments as well as
the theoretical expectations for $\bar p /p$-ratio are presented
in fig. 1.

An observation even of a single antinuclei in cosmic rays would be a
decisive evidence in favor of existence of a large amount of antimatter
in the Universe because the probability of their secondary production
is negligible. No compelling evidence indicating the existence of
antinuclei in the cosmic rays has been found though there are a few
strange events which might be interpreted this way (see ref. [22]).

$p \bar p$-annihilation into $\pi^0$'s with the subsequent decay of the
latter into $2\gamma$ may be observed in cosmic $\gamma$-rays at the
Earth. In particular the antimatter annihilation might explain the
observed isotropic cosmic gamma-ray background. Though this explanation
is rather speculative there is no other known  satisfactory mechanism
and the origin of this background radiation remains weird.
The flux of cosmic $\gamma$-rays from $p \bar p$-annihilation as well
as from neutralino annihilation was calculated in ref. [25]. Its
results are presented in fig. 2.

At the moment none of the observations can be considered as a confirmation
of the existence of any considerable amount of antimatter in the
Universe but they neither can reject the hypothesis. Evidently it can
never be rejected since the observations can only put a lower limit
on the distance to the antimatter-rich region. The flux of cosmic
$\bar p$ shows that this distance is larger than 10 Mpc.
\bigskip
\centerline {{\bf 6. Gravitational interaction of antimatter and new
long-range forces.}}
\bigskip

Long-range forces are known to be created by massless particle exchange
which can results in the potential $U(r) \sim 1/r$ and the force
$U'(r) \sim 1/r^2$. An exchange of massive particles gives rise to the
Yukawa type potential $U\sim \exp (-mr ) /r$. There are only two
known massless particles in Nature: the photon and the graviton. The latter
strictly speaking is not discovered experimentally but it is hard
to believe that gravitons do not exist. Zero mass of photons is maintained
by the gauge invariance of electromagnetism while that of gravitons is
maintained by the general covariance. If not for that quantum corrections
should definitely give rise to nonvanishing masses even if they were
zero at the classical level. Zero mass of both photon and graviton are
experimentally established with a very good accuracy. (One may ask
how it is possible to put a bound on the graviton mass if it
is not experimentally proven that the particle exists. The answer is
that if we assume that graviton exists then the data shows that
its mass should be very small
or exactly zero.) A good agreement of the Newton law at the scale of the
Solar system with astronomical data implies that the Compton wave length
of the graviton $\lambda _g \equiv m_g^{-1} $ is larger than the size of
the Solar system and the recent observations of the gravitational lenses
shows that $\lambda _g > 10$ Kpc. Analogously the observation of the
Jupiter magnetic field means that $\lambda _\gamma > 10^4 $km while the
existence of the galactic magnetic fields demands $\lambda _\gamma
>10$ Kpc.

A modification of gravitational interaction which breaks general
covariance should generically result in nonzero $m_g$ and so
theoretically disfavored. One can mimic a breaking of the equivalence
principle by introduction of new long-range forces and correspondingly
new massless particles. Introduction of new massless
tensor particles (that is those
with spin 2) is not possible because there is only one conserved
tensor quantity (energy-momentum tensor) which can be the source of
such particles and this source is already occupied by gravitons.

Long-range forces due to exchange of scalar particles
are not theoretically
supported. There is no symmetry which can maintain $m=0$ in the scalar
case. For this reason the Brans-Dicke modification of gravity would be
effective only at a short distance. The dilaton field associated
with conformal symmetry should be massive because the symmetry is known
to be badly broken. It is known that there should appear massless
scalar bosons if a global symmetry is spontaneously broken (the
so-called goldstone bosons) but it can be shown that the coupling of
these bosons to matter is of pseudoscalar form, like $\phi \bar \psi
\gamma_5 \psi $, so that their exchange gives rise to the force
falling as $1/r^3$.

The only remaining possibility is vector particle exchange. One can
invent a number of conserved vector currents (with the corresponding
gauge symmetry) which might be sources
for massless vector particles. In particular currents connected with
baryonic (B) and leptonic (L) charges are especially interesting.
However B and L are not conserved because of quantum chiral anomaly.
Still their sum is conserved and there may be a massless vector boson
coupled to the corresponding current. A high accuracy with which the
equivalence principle is fulfilled demands that the corresponding gauge
coupling constant is very small, $\alpha _{B+L} <10^{-43}$. For
comparison the electromagnetic coupling constant is $\alpha _{em}
\approx 10^{-2}$.

Thus one may expect that there may be new
long-range forces associated with
the exchange of new massless vector bosons. These forces should be
attractive between matter and antimatter and repulsive between the
same kind of matter in contrast to forces induced by tensor (graviton)
exchange which are always attractive.
\bigskip
\centerline {{\bf 7. Conclusions.}}
\bigskip

\item {1.} The Universe may be charge symmetric having an equal amount
of matter and antimatter.
\item {2.} The size of matter-antimatter domains is absolutely unknown
and can be as small as 10 Mpc or be larger than the present-day
horizon. In the last case no observation of antimatter is possible in
any foreseeable future.
\item {3.} A noticeable amount of uniformly distributed compact objects
made of antimatter in the background of the baryonic Universe is possible
theoretically and permitted by observations.
\item {4.} No evidence of cosmic antimatter is found by looking for
$\bar p$-flux.
\item {5.} Search of antinuclei in cosmic rays is of great importance
and any positive result would be an unambiguous proof of the
existence of antimatter in considerable amount.
\item {6.} Gamma-ray background might be an indication of cosmic antimatter.
\item {7.} Gravitational interaction of matter and antimatter are most
likely the same but new long-range forces associated with the exchange of
a new massless vector particle are not excluded.

\bigskip

\centerline {{\bf References.}}
\bigskip

\item {1.} G. Steigman, Ann. Rev. Astron. Astrophys. 14 (1976) 336.
\item {2.} F. W. Stecker, Nucl. Phys. B252 (1985) 25.
\item {3.} A. D. Sakharov, Pis'ma ZhETF. 5 (1967) 32.
\item {4.} A. D. Dolgov and Ya. B. Zeldovich, Rev. Mod. Phys. 53 (1981) 1.
\item {5.} E. Kolb and M. Turner, Ann. Rev. Nucl. Part. Sci. 33 (1983) 645.
\item {6.} A. D. Dolgov, Phys. Repts. (1992) to be published.
\item {7.} H. Georgi and S. L. Glashow, Phys. Rev. Lett. 32 (1974) 438.
\item {8.} G. t'Hooft, Phys. Rev. Lett. 37 (1976) 8; Phys. Rev. D11 (1976)
       3432.
\item {9.} V. A. Kuzmin, V. A. Rubakov, and M. E. Shaposhnikov, Phys. Lett.
      B155, (1987) 171.
\item {10.} I. Affleck and M. Dine, Nucl. Phys.  B249 (1985) 361.
\item {11.} A. D. Linde, Phys. Lett. B116, (1982) 335; A. Vilenkin and
      L. H. Ford, Phys. Rev. D26 (1982) 1231.
\item {12.} T. D. Lee, Phys. Rev. D8 (1973) 1226.
\item {13.} J. Yokoyama, H. Kodama, K. Sato, and N. Sato, Int. J. Mod. Phys.
      A2 (1987)  1808.; J. Yokoyama, H. Kodama, and K. Sato, Prog. Theor.
      Phys. 79 (1988) 800.
\item {14.} R. W. Brown and F. W. Stecker, Phys. Rev. Lett. 43 (1979) 315.
\item {15.} K. Sato, Phys. Lett. B99 (1981) 66.
\item {16.} Ya. B. Zeldovich, I. Yu. Kobzarev, and L. B. Okun, ZhETF 67
      (1974) 3.
\item {17.} A. D. Dolgov, A. F. Illarionov, I. D. Novikov,
      and N. S. Kardashev, ZhETF 94 (1987) 1; M. V. Chizhov and A. D. Dolgov,
      Nucl. Phys. B372 (1992) 521.
\item {18.} T. J. Broadhurst, R. S. Ellis, D. C. Koo, and A. S. Szalay,
      Nature 343 (1990) 726.
\item {19.} A. Dolgov and J. Silk CfPA-92-04.
\item {20.} G. F. Smoot et al, Astrophys. J. 396 (1992) L1.
\item {21.} K. A. Olive, D. N. Schramm, G. Steigman, and T. P. Walker,
      Phys. Lett. B236 (1990) 454.
\item {22.} S. P. Ahlen, P. B. Price, M. H. Salamon, and G. Tarle,
      Astrophys. J. 260 (1982) 20.
\item {23.} Y.-T. Gao, F. W. Stecker, M. Gleiser, and D. Cline,
       Astrophys. J. 361 (1990) L37;
       Y.-T. Gao and D. Cline, Mod. Phys. Lett. A6 (1991) 2669.
\item {24.} M. H. Salamon, Astrophys. J. 349 (1990) 78.
\item {25.} F. W. Stecker, Nucl. Phys. (Proc. Suppl.) 10B (1989) 93.

\vfill
\eject

\centerline {{\bf Figure captions. }}

{\bf Fig. 1.} Bounds on $\bar p /p $-ratio from ref. [24]. These results
are labeled LV, HV, and Total. Also presented results of other experiments
as well as theoretical predictions for different mechanisms of antiproton
production.

{\bf Fig. 2.} The gamma-ray background spectrum from
$p \bar p $-annihilation as is calculated in ref. [25]. Also presented are
the extrapolated X-ray background component (X), the galactic high
latitude cosmic ray produced background, and spectra from 15 GeV
photino and higgsino annihilation.
\end